\documentstyle[12pt,epsf]{article}

\def\slashchar#1{\setbox0=\hbox{$#1$}           
   \dimen0=\wd0                                 
   \setbox1=\hbox{/} \dimen1=\wd1               
   \ifdim\dimen0>\dimen1                        
      \rlap{\hbox to \dimen0{\hfil/\hfil}}      
      #1                                        
   \else                                        
      \rlap{\hbox to \dimen1{\hfil$#1$\hfil}}   
      /                                         
   \fi}                                         %
\def\half{\textstyle{1\over 2}}
\def\nn{\nonumber}

\begin{document}

\begin{titlepage}

{\hspace*{\fill} EFI-93-67 \\
 \hspace*{\fill} hep-ph/9408225 \\
 \hspace*{\fill} July 1994\\ }

\bigskip\bigskip

\begin{center}
      {\Large\bf Regularization in the gauged Nambu--Jona-Lasinio model\\}
\end{center}

\bigskip

\begin{center}
{\large Tony Gherghetta\footnote{e-mail: tonyg@yukawa.uchicago.edu}\\}
\medskip\medskip
{\it Enrico Fermi Institute and Department of Physics,\\University of
Chicago, Chicago, IL 60637\\}
\end{center}

\bigskip

\begin{abstract}
Various prescriptions employed for regulating gauged
Nambu--Jona-Lasinio type models such as the top-quark condensate model
are discussed. The use of dimensional regularization maintains gauge
invariance but destroys the quad\-ratic divergence in the gap
equation.  If instead a simple ultraviolet momentum cutoff is used to
regulate loop integrals, then gauge invariance is destroyed by a
quadratically divergent term as well as by ambiguities associated with
arbitrary routing of loop momenta. Finally it is shown that one can
use dispersion relations to regulate the top-quark condensate model.
This prescription maintains gauge invariance and does not depend on
arbitrary shifts in loop momenta.
\end{abstract}

\end{titlepage}

\section{Introduction}

In the past few years, following the realization that the top quark is
heavier than the gauge boson masses, there has been renewed interest
\cite{csate} in the original Nambu--Jona-Lasinio (NJL) model
\cite{njl} to provide a possible dynamical symmetry breaking mechanism
for the Standard Model.  In particular, new strong forces at a high
energy scale are assumed to cause the formation of ${\bar t}t$ bound
states (top condensation) and dynamically break the SU(2) $\times$
U(1) symmetry. This leads to an effective low energy theory which is
qualitatively equivalent to the Standard Model, but with a heavy top
quark playing a direct role in the symmetry breaking.

A minimal scheme implementing the NJL mechanism for the Standard Model
was given by Bardeen, Hill and Lindner, (BHL) \cite{bhl} who obtained
precise predictions for the top and Higgs masses. They argued that the
low energy effective Lagrangian in the fermionic bubble approximation
gives rise to the usual gauge coupling $\beta$-functions in the
Standard Model from fermion loops. By imposing the compositeness
conditions as ultraviolet boundary conditions on the renormalization
group flow and assuming the full one-loop $\beta$-functions of the
Standard Model, BHL found that typically the top quark mass is heavy,
$m_{top}\simeq 225 \,\rm GeV$ and $m_{Higgs} \simeq 1.1\, m_{top}$,
for composite scales ranging from $10^{15} \rm GeV \rightarrow 10^{19}
GeV$. These quantitative predictions are controlled by the infra-red
fixed point structure of the renormalization group equations \cite{hill}
and consequently a heavy top mass is a generic feature of this model.

In order to show that the gauged NJL model is qualitatively equivalent
to the Standard Model one must take care in choosing a consistent
regularization scheme. This is because the four-Fermi interaction in
the NJL model leads to quadratically divergent terms which can destroy
many of the desired qualitative features of the Standard Model, before
any quantitative predictions can be made.

In this work we discuss various regularization schemes one can
implement in the gauged NJL models such as the top-quark condensate
model \cite{bhl}. In these models the dynamics leads to a gap equation,
where unlike in the original BCS theory, there are quadratically
divergent terms. While these terms lead to fine tuning problems, they
are absolutely necessary for any consistent model and the choice of
regularization scheme must be made compatible with them.  In order to
make contact with the Standard Model, the vector gauge boson masses in
the NJL model must be induced in a locally gauge invariant manner.
One common scheme often employed is dimensional regularization and we
will discuss the consequences of implementing this scheme consistently
for the NJL model. In addition Willey \cite{willey} recently argued
that the NJL model suffers from inherent ambiguities arising from the
arbitrary routing of momenta through quadratically divergent fermion
bubble graphs. Willey employed an ultraviolet momentum cutoff for all
integrals and we will examine the consequences of these ambiguities
for the top-quark condensate model. Finally we will show that a
consistent prescription for regulating the top-quark condensate model
is to use dispersion relations. This prescription is gauge invariant
and avoids the problems of ambiguities arising from quadratic
divergences.

\section{The NJL model and dimensional regularization}

Let us begin with the minimal top-condensate model of Bardeen, Hill
and Lindner \cite{bhl}. The NJL Lagrangian for the Standard
Model at the scale $\Lambda$ is given by
\begin{equation}
	{\cal L}= {\cal L}_{kinetic} +~G ({\bar\Psi}_{L a} t_R^a)
                 ({\bar t}_{R b}\Psi^b_L) \qquad ;
	\qquad \Psi_L=\left(\matrix{t_L \cr b_L}\right)
\label{njllag}
\end{equation}
where a and b are colour indices and ${\cal L}_{kinetic}$ contains
only the gauge boson and fermion kinetic energy terms. Note that
(\ref{njllag}) is an effective Lagrangian of some renormalizable interaction
above the cutoff scale $\Lambda$ and that the four-fermion coupling constant,
$G$ must be positive to ensure an attractive interaction. The induced top
quark mass resulting from the four-fermion interaction in (\ref{njllag}) is
obtained by solving the gap equation to leading order in $N_{colour}$
\begin{equation}
	1=2 G N_c i\int {d^4 k\over (2\pi)^4} {1\over k^2-m_t^2}\, .
\label{gapint}
\end{equation}
Assuming that (\ref{njllag}) describes physics up to the scale $\Lambda$ the
integral (\ref{gapint}) can be regularized with an ultraviolet Euclidean
momentum cutoff $\Lambda$.
This leads to the gap equation
\begin{equation}
	1={G N_c \over 8\pi^2}\left(\Lambda^2-m_t^2 \ln
	{\Lambda^2\over m_t^2}\right)\, ,
\label{gapeqn}
\end{equation}
where we have only kept leading order terms. Notice that there is a
quadratically divergent term in the gap equation (\ref{gapeqn}) which is
nothing other than the reincarnation of the Higgs boson mass quadratic
divergence in the Standard Model and leads to the usual gauge-hierarchy
problem.

The low energy effective Lagrangian (\ref{njllag}) is clearly gauge
invariant because it only contains the gauge boson kinetic energy.
When the scalar ${\bar t} t$ condensate forms, the gauge bosons must
acquire mass in a $SU(2)\times U(1)$ gauge invariant manner because an
explicit mass term would break gauge invariance.  In the top-quark
condensate model the gauge boson mass arises from the fermion mass
dependence of the gauge boson self-energy, and so one cannot simply
discard all the non-gauge invariant terms in the self-energy. If one
employs dimensional regularization then it turns out that only the
fermion mass dependent non-gauge invariant terms survive and they
combine with terms from the Nambu-Goldstone mode contributions, so
that the gauge bosons become massive in a gauge invariant manner.
However the unfortunate feature of the dimensional regularization
scheme is that it also kills all quadratic divergences. This is
because in dimensional regularization
\begin{equation}
	\lim_{m^2\to 0} \int {d^D k\over k^2-m^2} =0\, .
\label{quadlim}
\end{equation}
While this is fine for the quadratically divergent non-gauge invariant
terms in the gauge boson self-energy, it is a disaster for the gap equation
because dimensionally regularizing the integral in (\ref{gapint}) destroys
the quadratic divergence in (\ref{gapeqn}). In particular the gap equation
without the quadratically divergent term is given by
\begin{equation}
	1=-{G N_c \over 8\pi^2} m_t^2 \left(\ln{\mu^2\over m_t^2} +1\right)\,,
\label{dreggap}
\end{equation}
where $\mu$ is the renormalization scale. Eq. (\ref{dreggap}) would
mean that the four-Fermi coupling $G<0$, corresponding to a repulsive
interaction. This would contradict the assumed attractive interaction
between top quarks in (\ref{njllag}). Clearly we need a regulator that
is sensitive to the quadratic divergences.

\section{Ambiguities in the NJL model}

\begin{figure}[b]
\vspace{0.2in}
\epsfxsize=\hsize
\epsffile{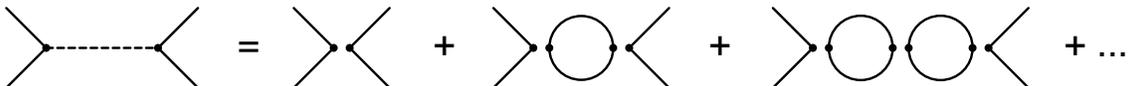}
\caption{\label{scalarfig} \small \it
	The infinite sum of fermion bubble diagrams. The scalar ${\bar t}t$
	channel has a pole at $2 m_t$ which corresponds to the composite
	Higgs scalar. The pseudoscalar ${\bar t}t$ and flavoured $tb$ channels
	give rise to three massless Nambu-Goldstone bosons.}
\end{figure}

The simplest prescription for regulating divergent integrals that does
not destroy the quadratic divergences in the gap equation is to use an
ultraviolet Euclidean momentum cutoff $\Lambda$. Let us now examine the
consequences of using this simple prescription for the top-condensate
model. Consider the sum of fermion-fermion scattering amplitudes
depicted in Figure~\ref{scalarfig}, where the amplitudes are summed to leading
order in $N_{colour}$ and the QCD coupling constant is zero (fermion
bubble approximation). Assuming that the gap equation is satisfied,
the fermion-fermion scattering amplitude in the scalar ${\bar t}t$
channel is given by
\begin{eqnarray}
\label{scalar}
	\Gamma_s(p^2)&=&{iG\over 2} \left(1-{iG\over 2}N_c \int
	{d^4 l\over (2\pi)^4} {\rm Tr}\left[{1\over \slashchar{l}
	+(1-\alpha)\slashchar{p} -m_t} {1\over \slashchar{l}-\alpha
	\slashchar{p} -m_t}
	\right] \right)^{-1} \nn \\
	&\simeq& {(4\pi)^2\over 2 N_c} \left({1\over 2}\Delta(\alpha)p^2
	+(p^2-4 m_t^2)\int_0^1 dx \ln{\Lambda^2\over m_t^2-p^2 x(1-x)}
	\right)^{-1}\, ,
\end{eqnarray}
where $\alpha$ is an arbitrary routing parameter introduced in the
\begin{figure}[b]
\vspace{0.2in}
\epsfxsize=0.5\hsize
\centerline{\epsffile{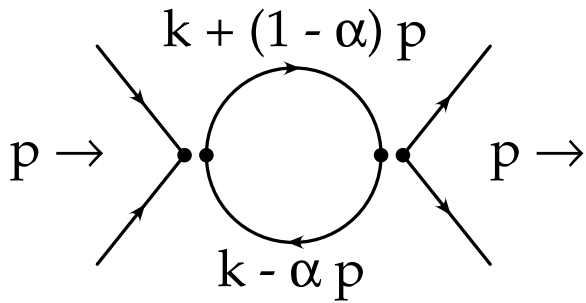}}
\caption{\label{routingfig} \small \it
	The loop momentum routing in a fermion bubble diagram.}
\end{figure}
loop momentum (see Figure~\ref{routingfig}) and $\Delta(\alpha)=
(1-2\alpha+2\alpha^2)$. Normally the parameter $\alpha$ is neglected
because the momentum integration variable can be shifted by a finite
amount for integrals which are at most logarithmically divergent.
However for integrals which are more divergent, this shift of
integration variable results in the appearance of an extra ``surface
term'' which depends on $\alpha$ \cite{jauch}. This causes the pole in
the scattering amplitude (\ref{scalar}) to shift by an amount
proportional to $\Delta(\alpha)/\ln(\Lambda^2/m_t^2)$ from the usual
value at $p^2=4\, m_t^2$. However since $\Lambda$ is necessarily
finite and $\alpha$ is an unrestricted constant, the pole receives an
arbitrary correction and so makes the relation between the
top and physical Higgs boson mass arbitrary. Note that when there is
no routing ambiguity or $\alpha=0$, the correction to the pole term is
negligible for large $\Lambda$.

This momentum routing ambiguity, in the context of the NJL model, was
first noticed by Willey \cite{willey} and is the result of using a momentum
cutoff regulator. It should be remarked that this routing ambiguity is
completely different from an earlier ambiguity noted by Hasenfratz et
al. \cite{hasenfratz}, who considered a lattice formulation of the NJL
theory.  In the lattice theory, higher derivative interaction terms
can be added to the NJL Lagrangian, which are in equivalent
universality classes, in the sense of defining the theory by going to
a critical point.  These terms introduce additional arbitrary
constants and give an ambiguous prediction for the ratio
$m_\sigma^2/m_f^2$. However assuming that the non-renormalizable
four-Fermi interaction arises from a renormalizable gauge theory at
high energy, it can be shown that the additional arbitrary constants
associated with higher derivative interaction terms are numbers of ${\cal
O}(1)$ and do not greatly influence the BHL predictions \cite{bardeen}.

Similarly in the neutral pseudoscalar ${\bar t}t$ channel the arbitrary
surface term, $\Delta(\alpha)$ will appear. The scattering amplitude is
given by
\begin{equation}
\label{pseudoscalar}
	\Gamma_P(p^2)\simeq {(4\pi)^2\over 2N_c p^2}\left({\Delta(\alpha)
	\over 2}+\int_0^1 dx~\ln{\Lambda^2\over m_t^2-p^2 x(1-x)}
	\right)^{-1}\, ,
\end{equation}
where the gap equation has again been invoked. Notice that the neutral
Nambu-Goldstone pole still occurs at $p=0$, except that now the coupling
constant depends on the arbitrary parameter $\alpha$. A similar dependence
on the arbitrary parameter $\alpha$ occurs for the charged Nambu-Goldstone
modes when one performs the fermion bubble sum in the flavoured $tb$
channel. Assuming $m_b\simeq 0$ one obtains for the scattering amplitude
in the $tb$ channel
\begin{equation}
\label{flavour}
	\Gamma_F(p^2)\simeq {(4\pi)^2\over 8N_c p^2}\left({\Delta(\alpha)
	\over4}+\int_0^1 dx~(1-x)\ln{\Lambda^2\over m_t^2(1-x) -p^2 x(1-x)}
	\right)^{-1}\, .
\end{equation}
Again notice that the routing ambiguity affects only the coupling constant
and not the masslessness of the charged Nambu-Goldstone boson.

\begin{figure}[b]
\vspace{0.2in}
\epsfxsize=\hsize
\epsffile{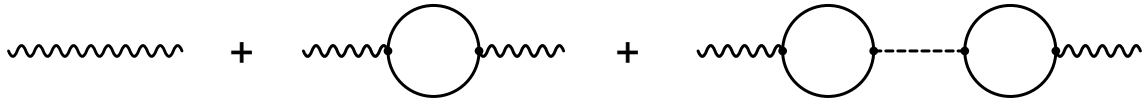}
\caption{\label{bosonfig} \small \it
	Diagrams contributing to the gauge boson vacuum polarization in
	the fermion bubble approximation.}
\end{figure}

The gauge bosons in the NJL model become massive by ``eating'' up the
dynamically generated Nambu-Goldstone modes. This corresponds to a
dynamical Higgs mechanism. Since there are routing ambiguities in the
Nambu-Goldstone modes it is clear that the gauge boson masses will
also be afflicted with this arbitrariness. The question of whether
gauge invariance can survive this arbitrariness needs to be checked.

Consider first the W-boson propagator, where the gauge fields are
rescaled so that the kinetic energy is $-1/(4 g^2) F^{\mu\nu}
F_{\mu\nu}$. In the fermion bubble approximation, corrections to the
propagator arise from the diagrams depicted in Figure~\ref{bosonfig}.
Again we will choose to regularize all divergent integrals with an
ultraviolet cutoff, $\Lambda$ and introduce an arbitrary parameter,
$\alpha$ to represent the routing ambiguity in the loop momentum.
The W-boson inverse propagator is given by the sum of three terms
\begin{equation}\label{wboson}
        {1\over g_2^2}D_{\mu\nu}^W(p)^{-1}={1\over g_2^2}(p_\mu p_\nu-
	g_{\mu\nu}p^2) +\Pi_{\mu\nu}(p) -{\textstyle{1\over 8}}
	J_\mu(p)\Gamma_F(p^2)J_\nu(p)\, ,
\end{equation}
where $g_2$ is the SU$(2)$ coupling constant and
\begin{equation}
\label{wbosontermsA}
	\Pi_{\mu\nu}(p)={i\over 8}\int {d^4 l\over (2\pi)^4} {\rm Tr}
	\left[\gamma_\mu(1-\gamma_5){1\over \slashchar{l}+(1-\alpha)
	\slashchar{p} -m_b} \gamma_\nu (1-\gamma_5) {1\over \slashchar{l}
	-\alpha \slashchar{p} -m_t}\right]\,,
\end{equation}
\begin{equation}
\label{wbosontermsB}
	J_\mu(p)=i\int {d^4 l\over (2\pi)^4} {\rm Tr} \left[\gamma_\mu
	(1-\gamma_5){1\over \slashchar{l}+(1-\alpha)\slashchar{p}-m_b}
	(1+\gamma_5) {1\over \slashchar{l}-\alpha \slashchar{p} -m_t}\right]\,.
\end{equation}
In Eq.(\ref{wboson}) $\Pi_{\mu\nu}$ is the usual one loop gauge boson
self-energy. There will be two types of divergent integrals appearing in
(\ref{wbosontermsA}) and (\ref{wbosontermsB}) which are not invariant under
an arbitrary shift of the loop momentum. These are given by
\begin{equation}
\label{integralA}
	\int {d^4 l\over (2\pi)^4}{l_\mu\over[(l+\chi p)^2-M^2]^2}
	=\int {d^4 l\over (2\pi)^4}{-\chi p_\mu\over(l^2-M^2)^2}
	+{i\over 32\pi^2} \chi p_\mu\,,
\end{equation}
\begin{eqnarray}
\label{integralB}
	\int {d^4 l\over (2\pi)^4}{l_\mu l_\nu\over[(l+\chi p)^2-M^2]^2}
	&=&\int {d^4 l\over (2\pi)^4}{(l_\mu l_\nu+\chi^2 p_\mu p_\nu)
	\over(l^2-M^2)^2} +{i\over 96\pi^2} \chi^2 (g_{\mu\nu} p^2-p_\mu
	p_\nu) \nn \\
	+{i\over 4\pi^2} \chi^4 p_\mu p_\nu &p^2&\int_0^1 dy \int_0^1 dz
	{y^3 z^2 \over M^2-\chi^2 p^2 y(1-y+y z(1-z))} \,,
\end{eqnarray}
where $\chi$ is an $l$ and $p$ independent constant. Using these expressions
and assuming the top quark mass satisfies the gap equation (\ref{gapeqn}) the
W-boson vacuum polarization becomes
\begin{equation}
\label{wbosonexp}
	{1\over g_2^2}D_{\mu\nu}^W(p)^{-1}=\left({p_\mu p_\nu\over p^2}
	-g_{\mu\nu}\right)
	\left[{p^2\over {\bar g}_2^2(p^2,\alpha)} -{\bar f}^2(p^2,\alpha)
	\right] -g_{\mu\nu} {\bar h}^2(p^2,\alpha)\,,
\end{equation}
where
\begin{equation}
\label{intdefnA}
	{1\over {\bar g}_2^2(p^2,\alpha)}={1\over g_2^2}+{N_c\over (4\pi)^2}
	\left[\int_0^1 dx~2x(1-x)\ln{\Lambda^2\over M^2(x)}
	-\textstyle{13\over 18}+\textstyle{5\over 3}\alpha (1-\alpha)
	-8 B(\alpha,M) \right]\,,
\end{equation}
\begin{equation}
\label{intdefnB}
	{\bar f}^2(p^2,\alpha)=m_t^2{N_c\over (4\pi)^2} {\left
	(\int_0^1 dx~(1-x)\ln{\Lambda^2\over M^2(x)}-\half(\half+\alpha)
	\right)^2 \over	\int_0^1 dx~(1-x)\ln{\Lambda^2\over M^2(x)}
	-\half(\half+\alpha-\alpha^2)}\,,
\end{equation}
\begin{equation}
\label{intdefnC}
	{\bar h}^2(p^2,\alpha)={N_c\over (4\pi)^2}\half\left[\Lambda^2
	-m_t^2 A(\alpha,M)+p^2({\textstyle{1\over 3}}-\alpha+\alpha^2 +
	8 B(\alpha,M))\right]\,,
\end{equation}
\begin{equation}
\label{intdefnD}
	A(\alpha,M)={\alpha(1+\alpha)\int_0^1 dx~(1-x) \ln
	{\Lambda^2\over M^2(x)}-{1\over 4}\alpha (1+3\alpha)\over
	\int_0^1 dx~(1-x) \ln{\Lambda^2\over M^2(x)}-\half(\half
	+\alpha-\alpha^2)}\,,
\end{equation}
\begin{equation}
\label{intdefnE}
	B(\alpha,M)=p^2\int_0^1 dx \int_0^1 dy \int_0^1 dz
	{(x-\alpha)^4 y^3 z^2 \over M^2(x)-(x-\alpha)^2 p^2 y(1-y+y z(1-z))}\,,
\end{equation}
and $M^2(x)=m_t^2(1-x)-p^2 x(1-x)$. The expression for the W-boson
inverse propagator (\ref{wbosonexp}) contains a quadratic divergence $(\propto
g_{\mu\nu} \Lambda^2)$ which does not appear in the corresponding
expression from BHL \cite{bhl}. This stems from the fact that an ultraviolet
cutoff has been used instead of dimensional regularization. This quadratic
term destroys gauge invariance and was first noticed a long time ago by
Wentzel \cite{wentzel}. Apart from this quadratically divergent term, which
would appear irrespective of loop momentum ambiguities, note that the
arbitrary constant $\alpha$ is enough by itself to destroy gauge invariance.
Unfortunately the ambiguity in the one loop W-boson self-energy and the
Nambu-Goldstone boson term do not combine together to save gauge invariance.

Similarly for the neutral gauge bosons we consider the contributions
arising from the diagrams in Figure~\ref{bosonfig} and include all surface
terms arising from divergent integrals. Working in the $(B-W^3)$ basis we
obtain in the limit $m_b\simeq 0$
\begin{eqnarray}
\label{neutral}
	{1\over g_i g_j} D^0_{\mu\nu}(p)^{-1}&=&\left({p_\mu p_\nu\over p^2}
	-g_{\mu\nu}\right)\left(\left[
	\matrix{{1\over g_1^2(p^2,\alpha)}& 0 \cr
                0&{1\over g_2^2(p^2,\alpha)}}
	\right]p^2 -\left[\matrix{1&-1 \cr -1&1}\right] f^2(p^2,\alpha)
	\right) \nn \\
	&&\quad - g_{\mu\nu}\left(\left[\matrix{{20\over 9}& 0 \cr 0&
	{4\over 3}}\right] Q_t + \left[\matrix{{2\over 9}& 0 \cr 0&
	{2\over 3}}\right] Q_b - \left[\matrix{1&-1 \cr -1&1}\right]
	h^2(p^2,\alpha)\right)
\end{eqnarray}
where $g_1$ is the U$(1)$ gauge coupling constant, $g_2$ is the
SU$(2)$ gauge coupling constant and
\begin{eqnarray}
\label{neutdefnA}
	{1\over g_1^2(p^2,\alpha)}={1\over g_1^2}&+&{N_c\over (4\pi)^2}
	\left[\int_0^1 dx~x(1-x)\left(\textstyle{20\over 9}\ln{\Lambda^2
	\over M_t^2(x)}+\textstyle{2\over 9}\ln{\Lambda^2\over M_b^2(x)}
	\right)\right. \nn \\
	&&\quad\qquad-\left.\textstyle{11\over 9}(\textstyle{13\over 18}-
	\textstyle{5\over 3}\alpha(1-\alpha))-\textstyle{80\over 9}
	B(\alpha,M_t)-\textstyle{8\over 9} B(\alpha,M_b)\right]\,,
\end{eqnarray}
\begin{eqnarray}
\label{neutdefnB}
	{1\over g_2^2(p^2,\alpha)}={1\over g_2^2}&+&{N_c\over (4\pi)^2}
	\left[\int_0^1 dx~x(1-x) \left(\textstyle{4\over 3}\ln{\Lambda^2
	\over M_t^2(x)}+\textstyle{2\over 3}\ln{\Lambda^2\over M_b^2(x)}
	\right)\right. \nn \\
	&&\quad\qquad-\left.\textstyle{13\over 18}+\textstyle{5\over 3}
	\alpha(1-\alpha)-\textstyle{16\over 3} B(\alpha,M_t)-\textstyle{8
	\over 3} B(\alpha,M_b)\right]\,,
\end{eqnarray}
\begin{eqnarray}
\label{moredefnA}
	f^2(p^2,\alpha)&=&{N_c\over (4\pi)^2}\left[{p^2\over 3}\left(
	\int_0^1 dx~x(1-x) \ln{M_b^2(x)\over M_t^2(x)}
	-4 B(\alpha,M_t)+4 B(\alpha,M_b)\right)\right. \nn \\
	&&\qquad\quad+\left.{1\over2} m_t^2 {(\int_0^1 dx~\ln
	{\Lambda^2\over M_t^2(x)}-1)^2\over \int_0^1 dx~\ln{\Lambda^2
	\over M_t^2(x)} +\half\Delta(\alpha)-1} \right]\,,
\end{eqnarray}
\begin{equation}
\label{moredefnB}
	Q_i(p^2,\alpha,m_i)={N_c\over (4\pi)^2}{1\over 4}\left[\Lambda^2
	-m_i^2+p^2(\textstyle{1\over 3}-\alpha+\alpha^2+8B(\alpha,M_i))
	\right]\,,
\end{equation}
\begin{eqnarray}
\label{moredefnC}
	h^2(p^2,\alpha)&=&{N_c\over (4\pi)^2}\left[{m_t^2\over 4}\left(
	{(\Delta(\alpha)-{1\over 3})\int_0^1 dx~\ln{\Lambda^2\over M_t^2(x)}
	+{1\over 3}-{7\over 6}\Delta(\alpha) \over \int_0^1 dx~\ln{\Lambda^2
	\over M_t^2(x)} +\half\Delta(\alpha)-1}\right)\right. \nn \\
	&&\qquad\quad +\left.{2\over 3}p^2 (B(\alpha,M_t)-B(\alpha,M_b)
	\right]\,,
\end{eqnarray}
with $M_t^2(x)=m_t^2-p^2 x(1-x)$ and $M_b^2(x)=m_b^2-p^2 x(1-x)$.
Again we note that gauge invariance is destroyed by ambiguities coming
from the surface terms of the linearly and quadratically divergent
integrals as well as from the Wentzel term. In the limit that all
surface terms are zero and ignoring the quadratic divergence arising
from the cutoff regulator, the above expressions reduce to the results
of BHL \cite{bhl}. Thus by regulating divergent loop integrals with an
ultraviolet momentum cutoff, one finds that the gauged NJL model is
plagued with the problems of a non gauge-invariant Higgs mechanism and
ambiguous quantitative predictions resulting from arbitrary loop momentum
routing. These problems are solely an artifact of using a simple
momentum cutoff.

\section{Dispersion relations}

An alternative prescription for regulating the top-quark condensate
model is to use dispersion relations. This prescription was previously
employed by Nambu and Jona-Lasinio \cite{njl} in their dynamical model of
elementary particles. In this prescription one first calculates the
imaginary part of an amplitude $\cal A$ by means of Cutkosky's rule
\cite{cutkosky} and then forms the complete amplitude by use of the
unsubtracted dispersion relation
\begin{equation}
\label{amp}
	{\cal A}(p^2)={1\over\pi}\int_L d\kappa^2{{\rm Im}{\cal A}(\kappa^2)
	\over \kappa^2-p^2-i\epsilon}\, ,
\end{equation}
where L is some cut along the real axis and the surface term at infinity
has been neglected by assumption. In general the high energy behaviour
of $\cal A$ makes the right-hand side of (\ref{amp}) divergent. This
ultraviolet divergence is regulated by replacing the upper limit of
integration by a cutoff scale $4\Lambda^2$, which represents the maximum
total energy squared. For energies greater than this scale $\rm Im{\cal A}=0$.

The first immediate consequence of the dispersion relation (\ref{amp}) is that
quadratically divergent amplitudes are no longer dependent on an arbitrary
parameter $\alpha$ resulting from loop momenta shifts. This is because
to calculate the imaginary part of a one loop amplitude, for example, the
intermediate particle states must be put on-shell according to Cutkosky's
rule. As a result the arbitrary dependence cancels out and quantitative
predictions will not be jeopardized.

In the NJL model the fermion mass satisfies a gap equation and so we need
to reformulate this self consistent condition using dispersion relations.
Consider first the sum of fermion bubbles in the
pseudoscalar ${\bar t}t$ channel $(\gamma_5)$. The scattering amplitude
is given by
\begin{equation}
\label{pseudoreln}
	\Gamma_P(p^2)={G\over 2}{1\over 1-J_P(p^2)}\, ,
\end{equation}
where
\begin{equation}
\label{pseudoamp}
	J_P(p^2)={G N_c\over (4\pi)^2}\int_{4 m_t^2}^{4\Lambda^2}\, d\kappa^2
        {\kappa^2\over \kappa^2-p^2-i\epsilon} \sqrt{1-{4 m_t^2 \over
	\kappa^2}}\, .
\end{equation}
Requiring $J_P(0)=1$ or
\begin{equation}
\label{polecond}
	1={G N_c\over (4\pi)^2}\int_{4 m_t^2}^{4\Lambda^2}\, d\kappa^2
	\sqrt{1-{4 m_t^2 \over \kappa^2}}\, ,
\end{equation}
leads to a pole at $p^2=0$, which is the massless neutral Nambu-Goldstone
mode. Evaluating the integral in (\ref{polecond}) to leading order gives
rise to the self consistent condition
\begin{equation}
\label{drgap}
	1={G N_c\over 8\pi^2}\left(2\Lambda^2-m_t^2 \ln
	{\Lambda^2\over m_t^2}\right).
\end{equation}
This is equivalent to the gap equation obtained earlier, (\ref{gapeqn}), by
noting that the ultraviolet Euclidean momentum cutoff $(\Lambda_E)$
used in sections 2 and 3 is related to the dispersion integral cutoff
by $\Lambda_E = {\sqrt 2}\,\Lambda$.

Similarly for the scattering amplitude in the scalar ${\bar t}t$ channel
we obtain
\begin{equation}
\label{screlan}
	\Gamma_S(p^2)=-{G\over 2}{1\over 1-J_S(p^2)}\,,
\end{equation}
where
\begin{equation}
\label{scalaramp}
        J_S(p^2)={G N_c\over (4\pi)^2}\int_{4 m_t^2}^{4\Lambda^2}\,
	d\kappa^2 {\kappa^2 -4 m_t^2\over \kappa^2-p^2-i\epsilon}
	\sqrt{1-{4 m_t^2 \over \kappa^2}}\,.
\end{equation}
If we invoke the condition, (\ref{polecond}) then $J_S(4 m_t^2)=1$ and
$\Gamma_S$ will have a pole at $p^2=4 m_t^2$. This is the dynamically
generated scalar bound state or Higgs mode.

The remaining flavoured $tb$ channels will similarly give rise to
\begin{equation}
\label{freln}
	\Gamma_F(p^2)=-{G\over 4}{1\over 1-J_F(p^2)}\,,
\end{equation}
where
\begin{equation}
\label{fdisp}
	J_F(p^2)={G N_c\over (4\pi)^2}\int_{m_t^2}^{4\Lambda^2}\, d\kappa^2
	{\kappa^2\over \kappa^2-p^2-i\epsilon} \left(1-{m_t^2 \over
	\kappa^2}\right)^2\,,
\end{equation}
and the bottom mass has been neglected $(m_b =0)$. Again invoking the
self-consistent condition, (\ref{polecond}) leads to the massless charged
Nambu-Goldstone modes $(J_F(0)=1)$. Thus in the dispersion relation
approach the gap equation is replaced by the condition requiring massless
Nambu-Goldstone modes.

In the absence of spontaneous symmetry breaking, on-shell quantities
are always gauge invariant. When the symmetry is spontaneously broken
we need to check that the gauge bosons receive their mass in a gauge
invariant manner. As depicted in Figure~\ref{bosonfig}, the gauge bosons
acquire their mass by absorbing the massless Nambu-Gold\-stone modes.
In particular, evaluating the W-boson vacuum polarization diagrams using
dispersion relations, one obtains for the W-boson inverse propagator
\begin{equation}
\label{wprop}
	{1\over g_2^2}D_{\mu\nu}^W(p)^{-1}=\left({p_\mu p_\nu \over p^2}-
        g_{\mu\nu}\right) \left({p^2\over {\bar g}_2^2(p^2)}-{\bar f}^2(p^2)
	\right)\,,
\end{equation}
where the dispersion integrals for ${\bar g}_2^2$ and ${\bar f}^2$ are
\begin{equation}
\label{fgdefnA}
	{1\over{\bar g}_2^2(p^2)}={1\over g_2^2}+{1\over 3}{N_c\over(4\pi)^2}
	\int_{m_t^2}^{4\Lambda^2}\, d\kappa^2
	{1\over \kappa^2-p^2-i\epsilon}\left(1+{2 m_t^2 \over\kappa^2}\right)
	\left(1-{m_t^2 \over\kappa^2}\right)^2\,,
\end{equation}
\begin{equation}
\label{fgdefnB}
	{\bar f}^2(p^2)= {m_t^2\over 2}{N_c\over(4\pi)^2}\int_{m_t^2}^
	{4\Lambda^2}\,d\kappa^2 {1\over \kappa^2-p^2-i\epsilon}
	\left(1-{m_t^2 \over\kappa^2}\right)^2\,.
\end{equation}
Note that the W-boson inverse propagator is transverse and corresponds
to a gauge invariant Higgs mechanism. The induced W-boson mass appears
as a pole in the propagator and is a solution of the equation
\begin{equation}
\label{wmasseqn}
	{M_W^2\over {\bar g}_2^2(M_W^2)}-{\bar f}^2(M_W^2)=0\,.
\end{equation}
To leading order in $\Lambda$ and assuming $M_W \ll m_t$, the W-boson mass
is given by
\begin{equation}
\label{wmass}
	M_W^2=g_2^2 {N_c\over(4\pi)^2}{m_t^2\over 2}\log{\Lambda^2
	\over m_t^2}\,,
\end{equation}
which is similar to the mass relation derived by BHL. As an aside we
note that (\ref{wmass}) is the same as the one loop correction to the W-boson
mass in the Standard Model with an elementary Higgs scalar. The effective top
Yukawa coupling, $f_t$ obtained from (\ref{screlan}) is
\begin{equation}
\label{yukawa}
	{1\over f_t^2}={N_c\over(4\pi)^2}\log{\Lambda^2\over m_t^2}\,,
\end{equation}
and this is easily seen to be compatible with (\ref{wmass}), where
$M_W^2=\half g_2^2 m_t^2/ f_t^2$.

Finally for completeness we give the dispersion integrals for the
neutral gauge boson propagators. Working in the $(B-W^3)$ basis leads
to
\begin{equation}
\label{neutprop}
	{1\over g_i g_j} D^0_{\mu\nu}(p)^{-1}=\left({p_\mu p_\nu\over p^2}-
        g_{\mu\nu}\right)\left(\left[\matrix{{1\over g_1^2(p^2)}& 0\cr
                               0&{1\over g_2^2(p^2)}} \right]p^2
                 -\left[\matrix{1&-1 \cr -1&1}\right] f^2(p^2)\right)\,,
\end{equation}
where
\begin{equation}
\label{neutpropdefnA}
	{1\over g_1^2(p^2)}={1\over g_1^2}+{\textstyle {20\over 9}}
	{\cal J}(m_t^2,p^2,\Lambda^2)
	+{\textstyle {2\over 9}}{\cal J}(m_b^2,p^2,\Lambda^2)\,,
\end{equation}
\begin{equation}
\label{neutpropdefnB}
	{1\over g_2^2(p^2)}={1\over g_2^2}+{\textstyle {4\over 3}}
	{\cal J}(m_t^2,p^2,\Lambda^2)
	+{\textstyle {2\over 3}}{\cal J}(m_b^2,p^2,\Lambda^2)\,,
\end{equation}
\begin{equation}
\label{neutpropdefnC}
	f^2(p^2)={\textstyle {1\over 3}}p^2{\cal J}(m_t^2,p^2,\Lambda^2)
	-{\textstyle {1\over 3}}p^2{\cal J}(m_b^2,p^2,\Lambda^2)
	+m_t^2\,{\cal K}(m_t^2,p^2,\Lambda^2)\,,
\end{equation}
and the dispersion integrals are given by
\begin{equation}
\label{disintA}
	{\cal J}(m^2,p^2,\Lambda^2)={1\over 6}{N_c\over(4\pi)^2}
	\int_{4 m^2}^{4\Lambda^2}\, d\kappa^2 {1\over \kappa^2-p^2-i\epsilon}
	\left(1+{2 m^2 \over\kappa^2}\right)\sqrt{1-{4m^2 \over\kappa^2}}\,,
\end{equation}
\begin{equation}
\label{disintB}
	{\cal K}(m^2,p^2,\Lambda^2)={1 \over 2}{N_c\over(4\pi)^2}
	\int_{4 m^2}^{4\Lambda^2}\, d\kappa^2
        {1\over \kappa^2-p^2-i\epsilon}\sqrt{1-{4m^2 \over\kappa^2}}\,.
\end{equation}
It is apparent from (\ref{neutprop}) that the neutral gauge boson will
become massive in a gauge invariant manner. Thus in the case of spontaneous
symmetry breaking Eqs. (\ref{wprop}) and (\ref{neutprop}) show that gauge
invariance is saved by the Nambu-Goldstone mechanism when dispersion
relations are used.

The high energy renormalization group running of the SU$(2)$ and
U$(1)$ gauge coupling constants can be obtained from the
dispersion integral expressions (\ref{fgdefnA}),(\ref{neutpropdefnA}) and
(\ref{neutpropdefnB}). Assuming that $p^2 \gg m_t^2$ one obtains
\begin{equation}
\label{gonerg}
	16\pi^2 p^2 {d\over d p^2} {1\over g_1^2(p^2)}= -{11\over 27}
	N_c\,,
\end{equation}
and
\begin{equation}
\label{gtworg}
	16\pi^2 p^2 {d\over d p^2} {1\over {\bar g}_2^2(p^2)}=
	16\pi^2 p^2 {d\over d p^2} {1\over g_2^2(p^2)}= -{N_c\over 3}\,.
\end{equation}
Note that the SU$(2)$ gauge coupling running obtained from the charged
gauge boson and neutral gauge boson expressions have identical
high energy running as expected.

Let us now compare Eqs.(\ref{gonerg}) and (\ref{gtworg}) with the usual
$\beta$-functions. The Standard Model one-loop $\beta$-functions for
the SU$(2)$ and U$(1)$ gauge coupling constants are given by
\begin{equation}
\label{smbetafnA}
	16\pi^2 {d g_1\over d t}=\left[{1\over 6} +{11\over 27}N_c n_q
	+n_l \right] g_1^3\,,
\end{equation}
\begin{equation}
\label{smbetafnB}
	16\pi^2 {d g_2\over d t}=\left[-{43\over 6} +{N_c\over 3}
	n_q +{1\over 3} n_l \right] g_2^3\,,
\end{equation}
where $n_q (n_l)$ are the number of quark (lepton) generations. One
can see that the single generation quark loop contributions to the
Standard Model $\beta$-functions agree with the high energy running
(Eqs.(\ref{gonerg}) and (\ref{gtworg})) obtained in the fermion bubble
approximation from the dispersion integral expressions for the gauge
coupling constants. This agreement further establishes the consistent
equivalence between the minimal top-condensate model and the Standard
Model when dispersion relations are used.

Clearly, in order to go beyond the fermion bubble approximation in the
top-condensate model one can utilize the full one-loop $\beta$-functions
of the Standard Model to obtain definite predictions for the
top and Higgs mass as originally advocated by Bardeen, Hill and
Lindner \cite{bhl}.

\section{Conclusion}

The choice of a consistent regulator in the gauged NJL model is
important in order to establish a qualitative equivalence with the
Standard Model. The use of an ultraviolet Euclidean momentum cutoff
$\Lambda$ to regulate divergent loop integrals is not a good
regularization prescription for the gauged NJL model. This is because
the prescription suffers from ambiguities associated with quadratic
divergences, which destroy the gauge invariance of the theory. The
dependence on the arbitrary parameter $\alpha$ also means that no
quantitative conclusions could be drawn from the model if this
regulator were used. In addition, the simple cutoff prescription
produces the well known quadratic divergence $(\propto
g_{\mu\nu}\Lambda^2)$ which destroys gauge invariance. An alternative
prescription would be to use dimensional regularization which would
eliminate this quadratically divergent term and give rise to a gauge
invariant Higgs mechanism.  However the quadratic divergences in the
gap equation would also be destroyed, leading to an inconsistent
theory.

A more suitable regularization prescription for the top-quark
condensate model that maintains gauge invariance without destroying
the quadratic terms is to use unsubtracted dispersion relations. In
this case the gap equation is equivalent to the condition requiring
the masslessness of the Nambu-Goldstone bosons.  Arbitrary shifts in
the loop momenta do not affect quantitative predictions because
intermediate particle states are put on-shell when the imaginary part
of an amplitude is calculated. Thus provided the gap equation is
satisfied, the dynamical Higgs mechanism is gauge invariant with no
corresponding routing ambiguities. If the high energy running of the
gauge coupling constants is obtained from the dispersion integral
expressions in the fermion bubble approximation then the usual fermion
loop contributions to the Standard Model $\beta$-functions are
obtained. This consistently shows that the top-quark condensate model
is equivalent to Standard Model when dispersion relations are used.
The BHL predictions for the top and Higgs mass are then obtained by
employing the full one-loop $\beta$-functions of the Standard Model
and imposing the compositeness conditions of the gauged NJL model.

\section*{Acknowledgments}

I would like to thank Professor Y.~Nambu for initially suggesting this
problem and for many helpful discussions and comments. I would also
like to acknowledge Dr. C.~Hill and Dr. W.~Bardeen for useful comments.
This work was supported in part by NSF contract PHY-91-23780.

\end{document}